\documentclass[sigplan,screen]{acmart}

\AtBeginDocument{%
  }

\acmBooktitle{The Future of Weak Memory,
  January 15, 2024}
\newcommand{\telechat}{T\'el\'echat}

\usepackage{float}
\usepackage{listings}
\lstdefinestyle{mystyle}{
    basicstyle=\ttfamily\footnotesize,
    breakatwhitespace=false,         
    breaklines=false,                 
    captionpos=b,                    
    keepspaces=true,                 
    numbers=left,                    
    numbersep=5pt,                  
    showspaces=false,                
    showstringspaces=false,
    showtabs=false,                  
    tabsize=2,
    numbers=none
}

\usepackage{tikz}
\usepackage{tikz-cd}
\usepackage{mathtools}

\begin{document}

\title{Weak Memory Demands Model-based Compiler Testing}

\author{Luke Geeson}
\email{luke.geeson@cs.ucl.ac.uk}
\affiliation{%
  \institution{University College London, UK}
  \country{}
}

\renewcommand{\shortauthors}{Luke Geeson}

\begin{abstract}
  A compiler bug arises if the behaviour of a compiled concurrent program, as allowed by its architecture memory model, is not a behaviour permitted by the source program under its source model. One might reasonably think that most compiler bugs have been found in the decade since the introduction of the C/C++ memory model. We observe that processor implementations are increasingly exploiting the behaviour of relaxed architecture models. As such, compiled programs may exhibit bugs not seen on older hardware. To account for this we require model-based compiler testing.

  While this observation is not surprising, its implications are broad. Compilers and their testing tools will need to be updated to follow hardware relaxations, concurrent test generators will need to be improved, and assumptions of prior work will need revisiting. We explore these ideas using a compiler toolchain bug we reported in LLVM.
\end{abstract}

\maketitle

\section{Introduction}

Consider the Message Passing litmus test in Fig.~\ref{exc}. The outcome \texttt{\{P1:r0=0; y=2\}} is forbidden by the C/C++ memory model~\cite{C11} (Fig.~\ref{out}). If Fig.~\ref{exc} is compiled using LLVM (Fig.~\ref{exc2}), the outcome \texttt{\{P1:r0=0; y=2\}} is allowed (Fig.~\ref{out2}) under the Arm AArch64 model~\cite{{Seal:2000:AAR:517257}}. We found~\cite{swp} this compiler bug after updates~\cite{rlxswp} to the \texttt{herd} memory model tool.

\begin{figure}
\centering
  \begin{tabular}{c}
\begin{lstlisting}[language=C, style=mystyle]
{ *x = 0; *y = 0; } // fixed initial state

#define relaxed memory_order_relaxed
#define release memory_order_release
#define acquire memory_order_acquire

P0 (atomic_int* y,atomic_int* x) {
  atomic_store_explicit(x,1,relaxed);
  atomic_thread_fence(release);
  atomic_store_explicit(y,1,relaxed);
}
P1 (atomic_int* y,atomic_int* x) {
  atomic_exchange_explicit(y,2,release);
  atomic_thread_fence(acquire);
  int r0 = atomic_load_explicit (x,relaxed);
}

exists (P1:r0=0 /\ y=2)
\end{lstlisting}
  \end{tabular}
  \caption{Message Passing Litmus Test. We reported~\cite{swp} this bug in LLVM's implementation of \texttt{atomic\_exchange}. We found this using the \telechat{} compiler testing tool.}
  \label{exc}
\end{figure}

The compiled counterpart of Fig.~\ref{exc} did not exhibit this outcome under the Arm AArch64~\cite{aarch64} model until the \texttt{herd} tool was updated~\cite{rlxswp} to reflect the intent of the architecture specification~\cite{Seal:2000:AAR:517257}. Likewise, older hardware that does not implement the relaxations allowed by the architecture may not exhibit the behaviour. Consequently compiler testing tools that depend on old hardware miss this bug. This motivates a need for compiler testing techniques that are parameterised over source and architecture models. In CGO 2024~\cite{geeson2023compiler}, we will present the \telechat{} tool that does exactly this. We have found several concurrency bugs using \telechat{}, numerous tooling bugs, and as far as we know it the first tool of its kind to be deployed in industry automated testing of a production compiler (It is the first to automatically test Arm Compiler). Compilers and their testing tools need to account for future relaxations, as they occur.

Fig.~\ref{exc} is not generated by today's litmus test generators. Fig.~\ref{exc} is a special kind of \textit{message passing} test that discards the read of the \texttt{atomic\_exchange} and checks \texttt{y} instead. Historically message passing checks whether the read of the \texttt{atomic\_exchange} can be reordered past the read into \texttt{P1:r0}. Test generators, such as \texttt{diy}~\cite{Alglave:2012:FWM:2205506.2205523}, currently implement the historic case. Bugs similar to Fig.~\ref{exc} have thus only been found by manual inspection, until now. We conclude that improvements to test generators are needed.

Interestingly, this bug disappears under direct observation. Using thread-local reads \texttt{int r1 = atomic\_exchange...} (\texttt{P1} is an \textit{Observer} - see \textsection B.2~\cite{Seal:2000:AAR:517257}) makes the bug disappear. The exchange is compiled to an Arm \texttt{SWP} instruction and the LLVM dead register optimisation~\cite{LLVMdrd} rewrites the \texttt{SWP} destination register to the zero register (\texttt{WZR}), since the read of the \texttt{exchange} is unused (Fig.~\ref{exc2}). Unfortunately instructions, whose destination register is a zero register (\texttt{WZR}/\texttt{XZR}), are not regarded as doing a read for the purpose of a acquire barrier (\textsection C3.2.12 of ~\cite{Seal:2000:AAR:517257}). LLVM has introduced a bug (Fig.~\ref{out2}). This optimisation was allowed, following an assumption~\cite{cmmtest} that ``optimisations affecting only the thread-local state cannot induce concurrency compiler bugs''. Fig.~\ref{exc} challenges this assumption with a new kind of Heisenbug. We conclude that assumptions surrounding compiler testing with models may need to be revisited in light of hardware relaxations.

\begin{figure}
\centering
  \begin{tabular}{c}
\begin{lstlisting}[language=C, style=mystyle]
{ P1:r0=0; [y]=1; }
{ P1:r0=1; [y]=1; }
{ P1:r0=1; [y]=2; }
\end{lstlisting}
  \end{tabular}
  \caption{Outcomes of running Fig.~\ref{exc} under the C/C++ memory model~\cite{C11}. Note the absense of \texttt{\{P1:r0=0; y=2\}}.}
  \label{out}
\end{figure}

\begin{figure}
\centering
  \begin{tabular}{c}
\begin{lstlisting}[language=C, style=mystyle]
AArch64 test

{ [P1_r0]=0;[x]=0;[y]=0;
  uint64_t %P0_x=x;uint64_t %P0_y=y;
  uint64_t %P1_P1_r0=P1_r0;
  uint64_t %P1_x=x;uint64_t %P1_y=y }

  P0                |  P1                    ;
   MOV W9,#1        |   MOV W9,#2            ;
   STR W9,[X%P0_x]  |   SWPL W9, WZR,[X%P1_y];
   DMB ISH          |   DMB ISHLD            ;
   STR W9,[X%P0_y]  |   LDR W8,[X%P1_x]      ;
                    |   STR W8,[X%P1_P1_r0]  ;
                    |                        ;


exists (P1_r0=0 /\ [y]=2)
\end{lstlisting}
  \end{tabular}
  \caption{Compiled version~\cite{swp} of Fig.~\ref{exc}. Note the destination register of the \texttt{SWPL} is the zero register (\texttt{WZR}). }
  \label{exc2}
\end{figure}

\begin{figure}
\centering
  \begin{tabular}{c}
\begin{lstlisting}[language=C, style=mystyle]
{ P1:r0=0; [y]=2; } bug!
{ P1:r0=0; [y]=1; }
{ P1:r0=1; [y]=1; }
{ P1:r0=1; [y]=2; }
\end{lstlisting}
  \end{tabular}
  \caption{Outcomes of running Fig.~\ref{exc2} under the Arm AArch64 model~\cite{aarch64}. Note the presence of \texttt{\{P1:r0=0; y=2\}}.}
  \label{out2}
\end{figure}

\begin{figure}[H]
\centering
  \begin{tabular}{c}
\begin{lstlisting}[language=C, style=mystyle]
AArch64 test

{ [P1_r0]=0;[x]=0;[y]=0;
  uint64_t %P0_x=x;uint64_t %P0_y=y;
  uint64_t %P1_P1_r0=P1_r0;
  uint64_t %P1_x=x;uint64_t %P1_y=y }

  P0                |  P1                    ;
   MOV W9,#1        |   MOV W9,#2            ;
   STR W9,[X%P0_x]  |   SWPL W9, W15,[X%P1_y];
   DMB ISH          |   DMB ISHLD            ;
   STR W9,[X%P0_y]  |   LDR W8,[X%P1_x]      ;
   RET              |   STR W8,[X%P1_P1_r0]  ;
                    |   RET                  ;


exists (P1_r0=0 /\ [y]=2)
\end{lstlisting}
  \end{tabular}
  \caption{Compiled version of Fig.~\ref{exc} (fixed). Note the destination register of the \texttt{SWPL} is \textit{not} the zero register \texttt{WZR}, but rather register \texttt{W15}. We recommended this fix in our bug report~\cite{swp}.}
  \label{exc3}
\end{figure}

\begin{figure}
\centering
  \begin{tabular}{c}
\begin{lstlisting}[language=C, style=mystyle]
{ P1:r0=0; [y]=1; }
{ P1:r0=1; [y]=1; }
{ P1:r0=1; [y]=2; }
\end{lstlisting}
  \end{tabular}
  \caption{Outcomes of running Fig.~\ref{exc3} under the Arm AArch64 model~\cite{aarch64}. Note the absence of \texttt{\{P1:r0=0; y=2\}}.}
  \label{out3}
\end{figure}

\section*{Acknowledgment}
We thank supervisors James Brotherston, Earl Barr, and Lee Smith. Ana Farinha, Richard Grisenthwaite, Alastair Donaldson, John Wickerson, Tyler Sorensen, Wilco Dijkstra, Tomas Matheson, Arm’s Compiler Teams and Arm Architecture \& Technology Group for their feedback and assistance. This work was supported by the Engineering and Physical Sciences Research Council [grant number EP/V519625/1]. The views of the authors expressed in this paper are not endorsed by Arm or any other company mentioned.

\bibliographystyle{acm}
\bibliography{paper}

\end{document}